\documentclass[aps,prl,twocolumn,longbibliography,floatfix,superscriptaddress]{revtex4-2}

\usepackage{amsmath, amssymb}
\usepackage{mathtools}
\usepackage{lipsum}
\usepackage{times}
\usepackage{graphicx}
\usepackage{wasysym}
\usepackage{bm}
\usepackage{hyperref}
\usepackage{xspace}
\usepackage{siunitx}
\usepackage{color}
\usepackage{braket}

\definecolor{myColor}{rgb}{0.02,0.12,0.3}
\definecolor{myciteColor}{rgb}{0.39,0.7,0.89}
\hypersetup{colorlinks=true, linkcolor=myColor, filecolor=myColor, urlcolor=myColor, citecolor=myColor, urlcolor=myColor}
\graphicspath{{./Figures/}}

\newcommand{\UD}{\ensuremath{U_{\text{D}}}}

\DeclareSIUnit{\nK}{\nano\kelvin}
\DeclareSIUnit{\aB}{\emph{a}_0}
\DeclareSIUnit{\G}{G}

\newcommand{\g}{\tilde{g}}

\newcommand{\oz}{\omega_{z}}
\newcommand{\oF}{\omega_{\text{F}}}

\newcommand{\kD}{k_{\text{D}}}
\newcommand{\kF}{k_{\text{F}}}
\newcommand{\kFv}{{\bf k}_{\text{F}}}
\newcommand{\kcf}{k_{\text{cf}}}

\definecolor{myColor}{rgb}{0.02,0.12,0.3}
\definecolor{myciteColor}{rgb}{0.39,0.7,0.89}
\definecolor{emphColor}{rgb}{0.1,0.1,0.8}
\definecolor{roughColor}{rgb}{0.7,0.1,0.1}
\definecolor{doneColor}{rgb}{0.0, 0.5, 0.0}

\begin{document}

\preprint{APS/123-QED}

\title{
Emergence of isotropy and dynamic scaling in 2D wave turbulence in a homogeneous Bose gas
}
\author{Maciej Ga{\l}ka}
\email[Corresponding author: ]{mg850@cam.ac.uk}
\author{Panagiotis Christodoulou}
\author{Martin Gazo}
\author{Andrey Karailiev}
\author{\\Nishant Dogra}
\affiliation{Cavendish Laboratory, University of Cambridge, J. J. Thomson Avenue, Cambridge CB3 0HE, United Kingdom}
\author{Julian Schmitt}
\affiliation{Cavendish Laboratory, University of Cambridge, J. J. Thomson Avenue, Cambridge CB3 0HE, United Kingdom}
\affiliation{Institut f{\"u}r Angewandte Physik, Universit{\"a}t Bonn, Wegelerstra{\ss}e 8, 53115 Bonn, Germany}
\author{Zoran Hadzibabic}
\affiliation{Cavendish Laboratory, University of Cambridge, J. J. Thomson Avenue, Cambridge CB3 0HE, United Kingdom}

\date{\today}

\begin{abstract}

We realise a turbulent cascade of wave excitations in a homogeneous 2D Bose gas, and probe on all relevant time and length scales how it builds up from small to large momenta, until the system reaches a steady state with matching energy injection and dissipation.  This all-scales view directly reveals the two theoretically expected cornerstones of turbulence formation -- the emergence of statistical momentum-space isotropy under anisotropic forcing, and the spatiotemporal scaling of the momentum distribution at times before any energy is dissipated.

\end{abstract}

\maketitle

Turbulence is a multi-scale phenomenon that is still not understood on a microscopic level, but is believed to generically feature cascades of excitations across different length scales~\cite{Richardson:1922,Kolmogorov:1941,Obukhov:1941}. 
When Richardson introduced the concept of a turbulent cascade a century ago~\cite{Richardson:1922}, he posited that energy injected into a fluid at a large length scale flows without loss through momentum space until it is dissipated at some small length scale. This is now known as a direct cascade, and the concept has been extended to an inverse one~\cite{Kraichnan:1967}, from small to large length scales. 

In a fully developed steady state, with matching energy injection (at one length scale) and dissipation (at another), turbulence is commonly manifested in stationary power-law spectra of system-dependent quantities like energy~\cite{Grant:1962} or enstrophy~\cite{Kraichnan:1967,Rutgers:1998}. Such spectra have been observed in a variety of contexts~\cite{Grant:1962,Ghashghaie:1996,Rutgers:1998,Hwang:2000,Sorriso:2007}, ranging from ocean waves~\cite{Hwang:2000} to financial markets~\cite{Ghashghaie:1996}. However, little is established experimentally about how such non-equilibrium steady states emerge, starting from an equilibrium system. 

In this Letter, we observe this emergence for a direct wave cascade in a homogeneous two-dimensional (2D) atomic Bose gas~\cite{Chomaz:2015,Navon:2021}.
Compared to earlier work on 3D wave turbulence in a homogeneous Bose gas~\cite{Navon:2016}, our geometry allows us to directly view the system on all relevant length scales, from the large one where we inject energy by an oscillating force to the small one where it is eventually dissipated. Our work also complements studies of the inverse energy cascade associated with vortex turbulence in 2D Bose gases~\cite{Johnstone:2019,Gauthier:2019}, and opens possibilities for further research, ranging from the interplay of wave and vortex turbulence~\cite{Kwon:2021} to quantum simulation of processes believed to have taken place in the early universe~\cite{Chatrchyan:2021}.

\begin{figure*}[t]
    \centering

 \includegraphics[width=1\textwidth]{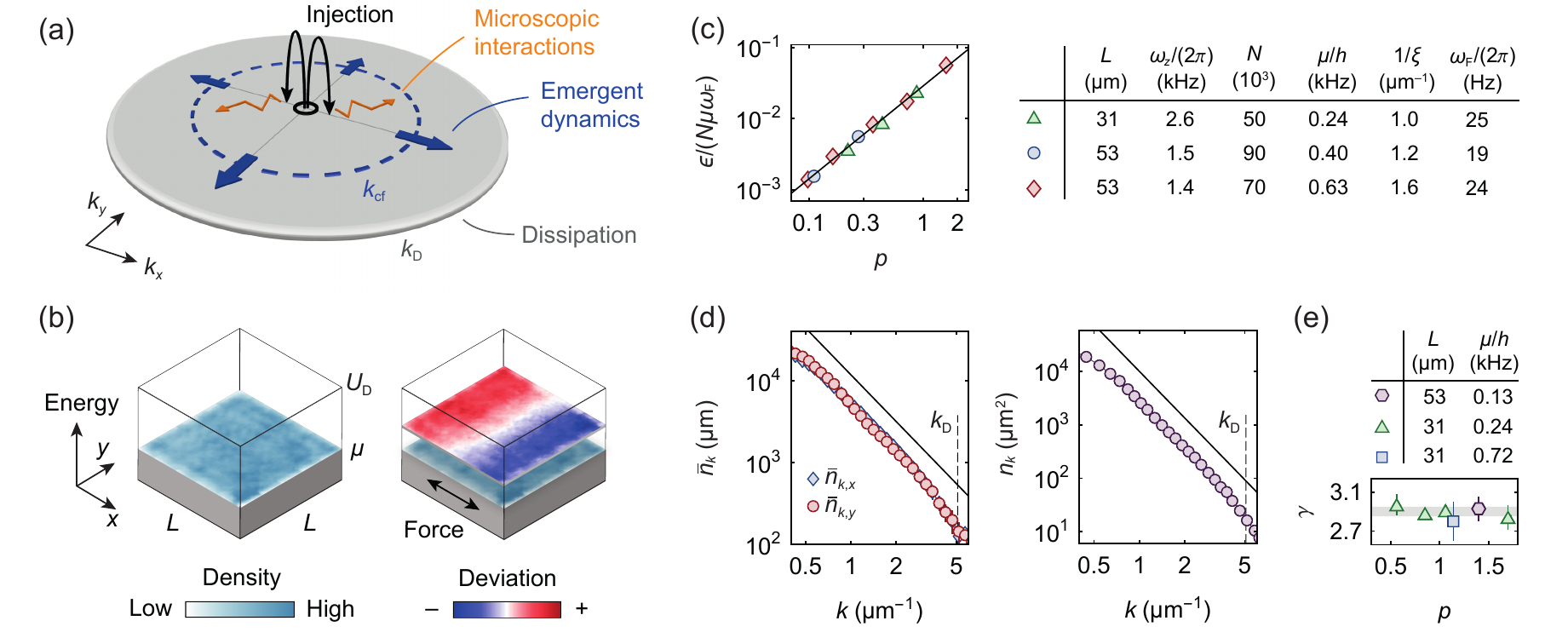}
\caption{
Direct wave cascade in a 2D quantum gas.
(a) Generic momentum-space picture of emergent isotropy and dynamic scaling. Anisotropic energy injection at a small wavenumber $k$ (black) and microscopic interactions (orange) lead to an isotropic cascade (blue), with the cascade front evolving as $\kcf \propto t^{-\beta}$ (where $t$ is time and $\beta<0$) until it reaches the dissipation scale $\kD$ and a steady state is established.
(b) Our experiment. Left: we start with a homogeneous 2D superfluid of $N$ atoms with chemical potential $\mu$ in a square box trap of size $L$ (here $53~\mathrm{\mu m}$) and depth $\UD$, which sets $\kD \propto \sqrt{\UD}$. Right: a spatially uniform force $F=F_0 \sin(\oF t)$ along $x$ resonantly injects energy into a phonon mode with wavevector ${\bf k}_{\rm F} = (\pi/L,0)$, visualised using PCA. 
(c)-(e) Universal steady-state features. 
(c)
The energy-injection rate, $\epsilon$, scaled by $N\mu\, \oF$ for different system parameters. The solid line shows $\epsilon/(N\mu\, \oF) \propto p^{1.31(3)}$, where $p=F_0L/\mu$. 
(d) Steady-state momentum distributions seen in TOF for $p=0.85$ and system parameters as for the green triangles in (c).
Left: line-integrated distributions parallel ($\bar{n}_{k,x}$) and perpendicular ($\bar{n}_{k,y}$) to the drive; right: azimuthally averaged $n_k(k)$. 
The solid lines show $n_k \propto k^{-\gamma}$ and $\bar{n}_{k,x(y)} \propto k^{-\gamma+1}$, with $\gamma=2.9$.
(e) Exponent $\gamma$ for different experimental parameters; shaded region shows $\gamma = 2.90(5)$.
}
 \label{fig:1}
\end{figure*}

The two key phenomena theoretically associated with the birth of direct-cascade wave turbulence are outlined in Fig.~\ref{fig:1}(a).
First, according to this picture, even though the continuous energy injection at a large length scale (small wavenumber $k = |{\bf k}|$)
is anisotropic, beyond a sufficiently large $k$ the cascade is statistically isotropic; such isotropy is believed to emerge in systems that, like our trapped gas, carry no net momentum~\cite{Falkovich:1993} (see~\cite{Monin:2013,Zakharov:1992}
for a more general discussion).
The second key phenomenon is dynamic scaling -- once the isotropic cascade front $\kcf$ forms, it evolves algebraically in time $t$, as $\kcf \propto t^{-\beta}$ (with $\beta<0$), until it reaches the dissipation scale $\kD$ and a steady state is established. In its wake, $\kcf$ leaves an isotropic power-law momentum distribution, $n_k ({\bf k}) \propto k^{-\gamma}$~\cite{Zakharov:1992,Nazarenko:2011}, so the pre-steady-state $n_k$ (at large $k$) follows the general form of self-similar spatiotemporal (dynamic) scaling:
\begin{equation}
(t/t_0)^{-\alpha} n_k \left(\mathbf{k},t\right)= n_k\left((t/t_0)^{\beta}\mathbf{k},t_{0}\right) \, ,
\label{eq:1}
\end{equation}
where $t_0$ is a reference time and in our case $\alpha = \gamma\beta$~\cite{Zakharov:1992,Nazarenko:2011}. Such scaling, known from classical surface growth~\cite{Family:1985,Kardar:1986}, and also seen in the relaxation dynamics of quantum gases~\cite{Pruefer:2018,Erne:2018,Glidden:2021,GarciaOrozco:2022, Wei:2022},
is hypothesised to be generic to far-from-equilibrium many-body quantum systems~\cite{Micha:2003,Berges:2008} and proposed as a way to classify them analogously to equilibrium universality classes.

As outlined in Fig.~\ref{fig:1}(b), we prepare a quasi-pure 2D superfluid of $^{39}$K atoms in a square optical box trap and drive it anisotropically with a time-periodic force created by a magnetic field gradient along one of the box axes, denoted $x$~\cite{Christodoulou:2021}. The gas is confined to the $x$-$y$ plane by a harmonic potential with angular trap frequency $\oz$, and has chemical potential $\mu= N \hbar^2 \g /(m L^2)$, where $N$ is the atom number, $L$ is the box size, $m$ is the atom mass, and $\g = \sqrt{8\pi m\oz /\hbar}\, a$~\cite{Petrov:2000a,Hadzibabic:2011}, where $a$ is the 3D $s$-wave scattering length, which we tune using a Feshbach resonance; for our typical system parameters see the legend in Fig.~\ref{fig:1}(c).
The spatially-uniform driving force, $F=F_0 \sin(\oF t)$, with $\oF =  c \,\pi/L$, where $c =\sqrt{\mu/m}$ is the speed of sound, resonantly injects energy into a longest-wavelength phonon mode, with wavevector $\kFv = (\pi/L, 0)$. Our energy-injection scale is thus set by the system size, $\kF = \pi/L \lesssim 0.1~\mu{\rm m}^{-1}$, while the dissipation scale $\kD =\sqrt{2m\UD/\hbar^2} \approx 5~\mu{\rm m}^{-1}$ is set by the trap depth $\UD$; the energy is dissipated from the system by particles with energy larger than $\UD$ leaving the trap~\cite{Navon:2019,Note1}. Between $\kF$ and $\kD$, the nature of excitations changes from phonons to particle-like matter waves at $k= 1/\xi$, where $\xi = L/\sqrt{\g N} \sim 1~\mathrm{\mu m}$ is the 2D healing length~\cite{Hadzibabic:2011}.

\begin{figure*}[t]
\centering
\includegraphics[width=\textwidth]{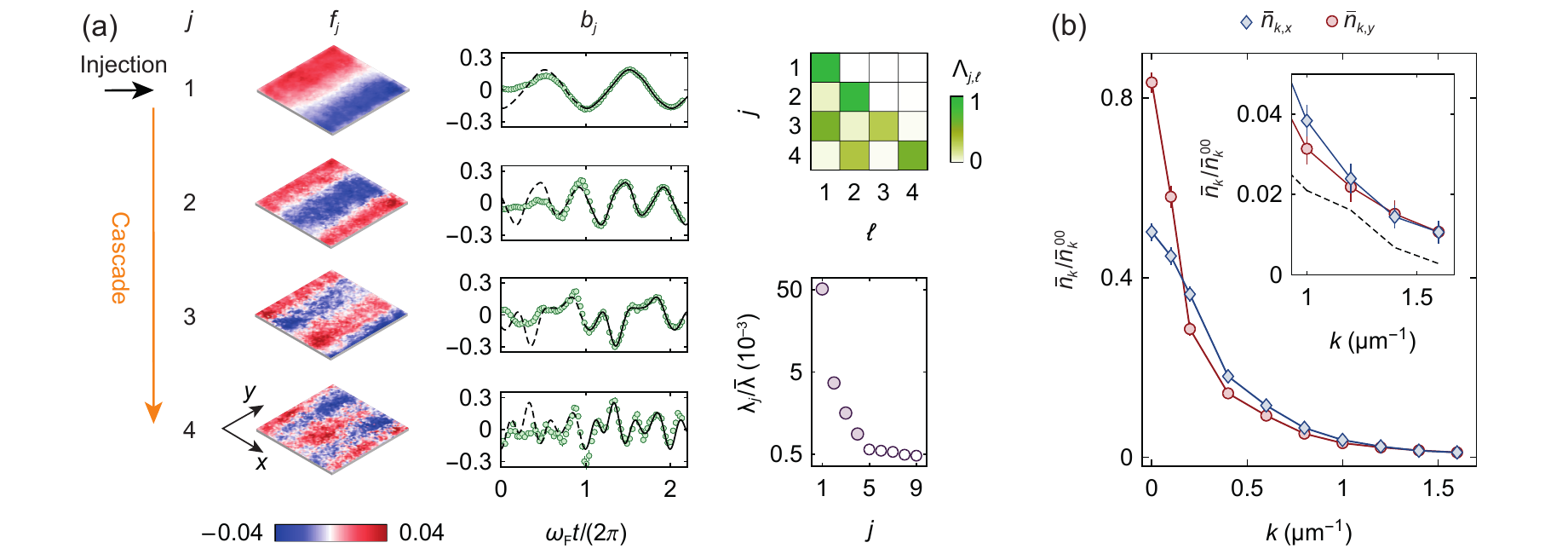}

\caption{
From low-$k$ anisotropy to high-$k$ isotropy.
(a) PCA decomposition of the {\it in-situ} density modulations, for $p=0.6$. The spatial structures, $f_j(x,y)$, of the first four modes show excitations only along the drive direction $x$. The temporal fits, $b_j(t) = \sum_{\ell=1}^4 B_{j,\ell} \cos(\ell \oF t+\phi_{j,\ell})$ for $t> 2\pi/\oF$ (solid lines, with the dashed ones showing extrapolations to shorter $t$), give the harmonic weights $\Lambda_{j,\ell} = B_{j,\ell}^2/(\sum_{\ell=1}^4B_{j,\ell}^2)$. The normalised PCA eigenvalues, $\lambda_j/\bar{\lambda}$, are also shown for the next five modes (open circles), which do not show any clear structures. 
(b) Emergence of isotropy seen in Bragg spectroscopy, for $p=0.85$ and $t=2\pi/\oF$. Here $\bar{n}_{k,x} (k)$ and $\bar{n}_{k,y} (k)$ are normalised by their common value, $\bar{n}_k^{00}$, measured for $k=0$ and $t=0$. The emergence of isotropy is seen in the convergence of the two curves for $k\gtrsim 1/\xi = 1.0~\mathrm{\mu m}^{-1}$; in the inset (zoomed-in at high $k$) the dashed line indicates distributions measured with similar error bars for $p=0$.
} 
\label{fig:2}
\end{figure*}

\begin{figure*}[t]
    \centering
    \includegraphics[width=\textwidth]{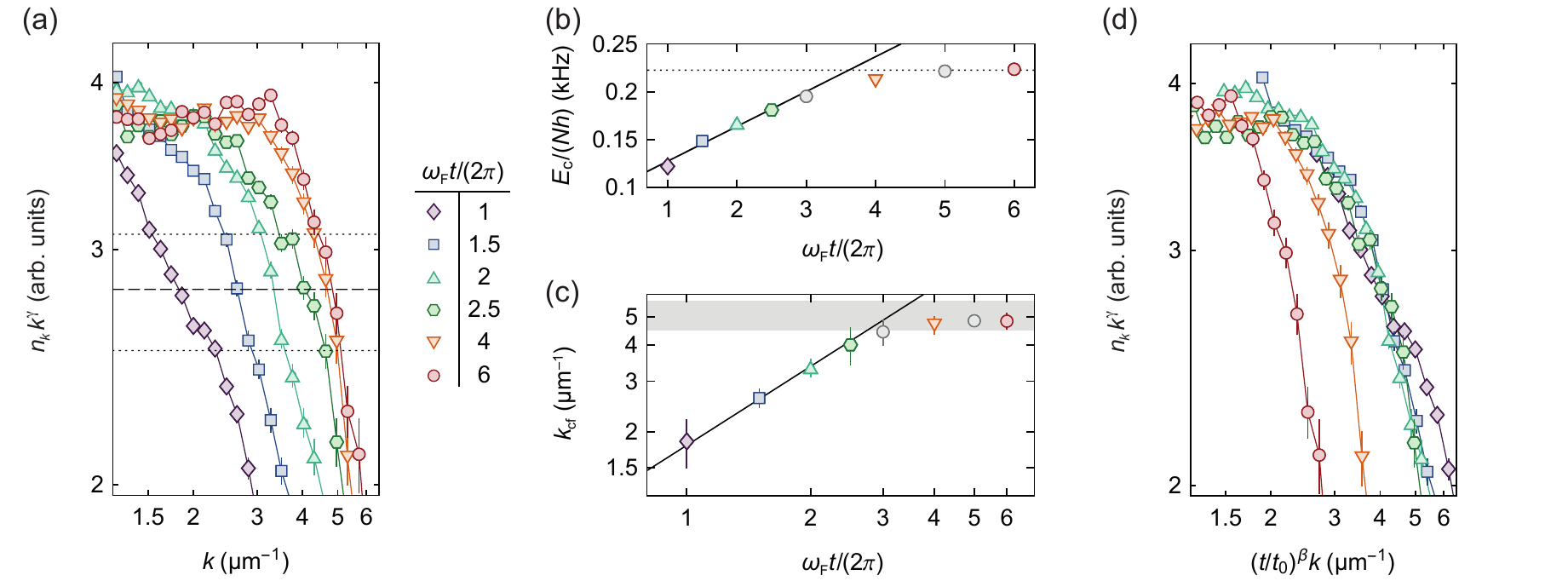}
  \caption{
Dynamic scaling in the pre-steady state. (a) Compensated spectra, $n_k k^\gamma$, with $\gamma=2.9$; note that $1/\xi = 1.0~\mathrm{\mu m}^{-1}$. (b) Total kinetic energy in the isotropic cascade, $E_c(t)$, obtained by integrating over the spectra in (a). The slope of the solid line is equal to the independently measured flux $\epsilon$ injected at $\kFv$, and the horizontal dotted line is a guide to the eye. For $t \gtrsim 3.5 \times 2\pi/\oF$ the steady-state cascade is established. (c) The cascade front, $\kcf (t)$, defined in (a) by the intersections of the data with the horizontal dashed line; error bars show systematic uncertainties defined by the intersections with the two dotted lines. The solid line shows the prediction  $\kcf \propto t^{-\beta}$, with $\beta=-0.91$ based on Eq.~(\ref{eq:beta}). A fit for $t< 3\times 2\pi/\oF$ (not shown) gives a consistent $\beta = -0.85(7)$. The shaded region corresponds to the independently estimated $\kD \propto \sqrt{\UD}$, including its uncertainty. (d) Compensated spectra from (a) rescaled according to Eq.~(\ref{eq:1}), with $\beta = -0.85$, $\alpha = \gamma \beta = -2.47$, and the arbitrary $t_0$ set to $2.5 \times 2\pi/\oF$. Note that $(t/t_0)^{-\alpha} \left((t/t_0)^{\beta} k \right)^{\gamma} n_k = n_k k^{\gamma}$, so when rescaling compensated spectra the $y$ axis remains the same. The collapse of the curves for $t<3\times 2\pi/\oF$ shows the dynamic scaling in the pre-steady state.
}
\label{fig:3}
\end{figure*}

We first summarise the universal features of our energy injection (Fig.~\ref{fig:1}(c)) and the resulting steady-state turbulence (Fig.~\ref{fig:1}(d), (e)), and then study how such a steady state gets established (Fig.~\ref{fig:2}, Fig.~\ref{fig:3}). To probe the gas on all length scales from $\kF$ to $\kD$, we use three complementary tools -- (i) using principal component analysis (PCA)~\cite{Segal:2010,Dubessy:2014} we directly visualise the dynamics of the low-lying ($k\sim\kF$) discrete quantum states, (ii) with time-of-flight (TOF) expansion we study the emergent statistical behaviour at large $k$, and (iii) using Bragg spectroscopy~\cite{Kozuma:1999a,Stenger:1999b} we bridge the $k$-space gap between these measurements.

To measure the energy flux $\epsilon$ injected at $\kFv$ (averaged over half a drive period), we monitor the periodic displacement of the cloud's centre of mass (COM), which is proportional to the density modulation due to the phonon excitation at $\kFv$ (see Fig.~\ref{fig:1}(b))~\cite{Christodoulou:2021}. Specifically, $\epsilon = N F_0 v_0/2$, where $v_0$ is the amplitude of the COM speed. Defining the dimensionless flux $\epsilon/(N\mu\, \oF) = p v_0/(2\pi c)$, where $p= F_0 L/\mu$ is the dimensionless drive strength~\cite{Navon:2016}, we find that it follows a universal curve, $\propto p^{1.31(3)}$ (Fig.~\ref{fig:1}(c)).
This scaling is in contrast with linear response (where $v_0 \propto F_0$, so $\epsilon \propto p^2$) and agrees with $\epsilon \propto p^{4/3}$ for a nonlinear transfer of energy to higher-lying excitations, as previously observed in 3D for a single interaction strength~\cite{Zhang:2021}.

For sufficiently strong drives, $p\gtrsim 0.5$ (corresponding to $v_0/c \gtrsim 0.15$), and at sufficiently long times, in TOF we observe steady-state power-law distributions such as  those shown in Fig.~\ref{fig:1}(d) for $p=0.85$ and $t = 5 \times 2\pi/\oF$. The line-integrated distributions parallel and perpendicular to the drive, $\bar{n}_{k,{x(y)}}(k_{x(y)}) = \int {\rm d}k_{y(x)} n_k(k_x, k_y)$, are essentially identical, implying an isotropic $n_k$. Note, however, that due to finite-size effects these measurements are not accurate for $k\lesssim 0.6~\mathrm{\mu m}^{-1}$~\cite{SI}. We also show the (azimuthally averaged) radial distribution $n_k(k)$, from which we extract $\gamma \approx 2.9$ (solid line). As shown in Fig.~\ref{fig:1}(e), $\gamma$ is robust under changes of the system parameters, including the box size, and the drive strength; from different measurements (always fitting in the range $1.5-3~\mathrm{\mu m}^{-1}$) we get a combined estimate $\gamma=2.90(5)$.

To trace how such a steady state gets established, we start with the onset of the cascade at low $k$, by studying {\it in-situ} the spatiotemporal modulations of the gas density, $n$ (Fig.~\ref{fig:2}(a)); here we use our larger, $53$-$\mathrm{\mu m}$ box, with parameters as for the red diamonds in Fig.~\ref{fig:1}(c) and $p=0.6$, while below for Bragg and TOF measurements we use a $31$-$\mathrm{\mu m}$ box with all parameters as in Fig.~\ref{fig:1}(d)~\cite{SI}. Using PCA, we decompose $n(x,y, t)$ in an unbiased way as $\sqrt{\bar{\lambda}}\, \bar{f} (x,y) + \sum_{j=1}^{J-1} \sqrt{\lambda_j} \, f_j(x,y) \, b_j(t)$,  with orthonormal $\{f_j(x,y)\}$ and $\{b_j(t)\}$, and $J$ equal to the number of different times for which we measure $n$. Here $\bar{f}$ is the normalised time-averaged density profile and $f_j$ are the principal components of the modulations $\Delta n(x,y,t)$, with eigenvalues $\lambda_j$ decreasing with increasing $j$, and $\sum_{j=1}^{J-1} \lambda_j/\bar{\lambda} =  \langle (\Delta n)^2 \rangle / \langle n \rangle^2$, where $\langle ... \rangle$ denotes an average over both space and time. For weak modulations, $f_{j}$ directly visualise the wavefunctions of the underlying excitations through interference with the quasi-uniform condensate.

We find that the first four $f_j$ (see Fig.~\ref{fig:2}(a)), with $f_1$ showing the resonantly excited phonon, all closely resemble phonon wavefunctions with ${\bf k} = j\kFv$; here $J=81$, but the first four modes account for $75\%$ of the total (normalised) density variance $\sum_{j=1}^{80} \lambda_j/\bar{\lambda} = 0.08$, and we do not identify any clear structures in the remaining ones. The directly driven $b_1$ oscillation at $\oF$ quickly reaches a steady state, while $b_2$ oscillates predominantly at $2\,\oF$ and with a discernible delay; $b_{3}$ and $b_{4}$ show more complex behaviour, but for $t> 2\pi/\oF$, all four $b_j$ are fitted well by $\sum_{\ell=1}^4 B_{j,\ell} \cos(\ell \oF t+\phi_{j,\ell})$, which gives their harmonic weights $\Lambda_{j,\ell} = B_{j,\ell}^2/(\sum_{\ell=1}^4B_{j,\ell}^2)$.
The nonlinear cascade naturally results in the appearance of the diagonal terms $\Lambda_{j,j}$, corresponding to $j\kFv$ phonons being created and revealed through interference with the condensate.
The prominent off-diagonal ones, $\Lambda_{3,1}$ and $\Lambda_{4,2}$, can be partially explained by noting that two-phonon interference also contributes to $\Delta n$ ({\it e.g.}, $B_{4,2}$ arising from interference of $\kFv$ and $3\,\kFv$ phonons); another contribution to $B_{3,1}$ arises from weak off-resonant direct driving of the $3\,\kFv$ phonon.

Crucially, up to $4\,\kF$, corresponding to $\approx 0.15/\xi$, all the dynamics are essentially one-dimensional. In the presence of a condensate, which makes a three-wave interaction (two phonons combining into a single higher-energy one) the dominant nonlinear process, such absence of cross-directional coupling is indeed theoretically expected for $k \ll 1/\xi$~\cite{Dyachenko:1992}.

To follow the fate of the anisotropy at higher $k$, we use Bragg spectroscopy, which gives the line-integrated distributions $\bar{n}_{k, x}$ and $\bar{n}_{k, y}$ without any finite-size artefacts. Normalising $\bar{n}_{k,x}(k)$ and $\bar{n}_{k,y}(k)$ to unity for zero $k$ and $t$~\cite{SI}, in Fig.~\ref{fig:2}(b) we show them for $t=2\pi/\oF$. By this time the excitations already cascade to $k > 1/\xi$ and, while at low $k$ their distribution is clearly anisotropic, at $k\gtrsim 1/\xi = 1.0~\mathrm{\mu m}^{-1}$ it is isotropic~\cite{Note2}.

With the isotropy of the momentum distribution established for $k\gtrsim 1/\xi$ and $t\ge 2\pi/\oF$, we turn to TOF measurements to study how $n_k(k,t)$ evolves until the steady state is reached.
In Fig.~\ref{fig:3}(a) we show the evolution of the compensated spectrum, $n_k k^{\gamma}$, which highlights the propagation of its leading edge. 
From $n_k$, measured at half-periods of the drive, we extract the total kinetic energy in the isotropic cascade ($k>1/\xi)$, $E_c(t) = \int {\rm d}k \, 2\pi k \, \varepsilon_k $, where $\varepsilon_k = n_k \, \hbar^2 k^2/(2m)$~\cite{SI} (Fig.~\ref{fig:3}(b)), and the cascade front $\kcf(t)$ (Fig.~\ref{fig:3}(c)).

Beyond some time, $t^* \approx 3.5\times 2\pi/\oF$, both $E_c$ and $\kcf$ saturate, as expected for a steady state with matching energy injection and dissipation~\cite{Note1}. Prior to that, the growth of $E_c$ is consistent with the independently measured $\epsilon$ injected at $\kFv$ (solid line); note that the systematic error in $\epsilon$ is $\approx 20\%$, dominated by the errors in the calibration of $N$ and $L$~\cite{SI}.
For $\kcf > 1/\xi$, associating a constant increase of $E_c$ with the growth of $\kcf$ leads to the scaling prediction~\cite{Zakharov:1992,Navon:2019} $\kcf \propto t^{-\beta}$ (for $t<t^*$), with
\begin{equation}
\beta =-1/(d+2-\gamma) \, ,
\label{eq:beta}
\end{equation}
where $d$ is the system dimensionality~\cite{Note3}. Note that this relation assumes a quadratic wave dispersion, which is our case for $k>1/\xi$~\cite{Note4}. For a quadratic dispersion, the analytical theory of weak-wave turbulence~\cite{Zakharov:1992,Nazarenko:2011} predicts $\gamma =d =2$ and $\beta = -1/2$, assuming very weak interactions and $\ln(\kD/\kF) \gg 1$. Our experimental $\gamma$ is different, and the origin of this difference remains to be elucidated (note that in a 3D gas $\gamma \approx 3.5$ was observed~\cite{Navon:2016}).
However, the relationship between $\beta$ and $\gamma$ in Eq.~(\ref{eq:beta}), which embodies the concept of dynamic scaling, should hold more generally~\cite{Navon:2019}, as its derivation is valid for any $\gamma < d+2$~\cite{Zakharov:1992,Nazarenko:2011}.
Taking our experimental $\gamma = 2.90(5)$ and $d=2$, we predict $\beta = -0.91(4)$, and in Fig.~\ref{fig:3}(c) show that $\kcf(t)$ agrees with this prediction (solid line). Alternatively, fitting $\kcf \propto t^{-\beta}$ for $t < 3\times 2\pi/\oF$ gives a consistent $\beta=-0.85(7)$ (not shown), with the error dominated by the systematic uncertainty in $\kcf (t)$.  

In Fig.~\ref{fig:3}(d) we show the data from Fig.~\ref{fig:3}(a) rescaled according to Eq.~(\ref{eq:1}).
The collapse of the curves for $t < t^*$ confirms the dynamic scaling in the pre-steady state, and we also show its breakdown at longer times. In the dynamics of closed quantum systems, such breakdown is expected when a system approaches equilibrium~\cite{Micha:2003,Berges:2008,Glidden:2021}; here it occurs when our driven gas reaches a non-thermal steady state.

In conclusion, our experiments provide a complete, all-scales picture of the birth of 2D wave turbulence, and our microscopic view on the far-from-equilibrium dynamics could allow many further studies. It would be interesting to vary the energy-injection scale, explore excitations above an established turbulent steady state, and study decaying turbulence~\cite{Nazarenko:2011}. In a broader context, such studies could also allow quantum simulation of the post-inflationary cosmological reheating~\cite{Chatrchyan:2021}. One could also search for scenarios in which the emergence of isotropy breaks down, for example by forcing the gas through a channel between two reservoirs~\cite{Brantut:2012}, so that turbulence forms in a moving frame. 

The supporting data for this Letter are available in the Apollo repository~\cite{Galka:2022data}.

\vspace{0.5cm}

We thank Lena Dogra, Christoph Eigen, Nir Navon, Giacomo Roati, and Henning Moritz for discussions and comments on the manuscript. This work was supported by EPSRC [Grants No.~EP/N011759/1 and No.~EP/P009565/1], ERC (QBox and UniFlat), STFC [Grant No.~ST/T006056/1] and QuantERA (NAQUAS, EPSRC Grant No.~EP/R043396/1). J.S. acknowledges support by the DFG [Grants No. 277625399 and EXC 2004/1–390534769] and from Churchill College (Cambridge). Z.H. acknowledges support from the Royal Society Wolfson Fellowship.

\clearpage
\setcounter{figure}{0} 
\setcounter{equation}{0}
\section{Supplemental material}
\subsection{Experimental system}
Our uniform two-dimensional gas of $^{39}$K atoms in the $\ket{F,m_F}=\ket{1,1}$ hyperfine ground state is optically confined in a potential sculpted using two Digital Micromirror Devices and calibrated as in~\cite{Christodoulou:2021}. Our atom number $N$ is calibrated with a systematic uncertainty of 15\% using measurements of the critical temperature for Bose--Einstein condensation in a 3D harmonic trap~\cite{Tammuz:2011}, and our box size $L$ has a systematic uncertainty of $1.5~\mu\mathrm{m}$. The systematic uncertainties in $\oz$ and $\g$ are $\lesssim 4\%$. The trap depth $\UD$ is set by the in-plane potential and experimentally estimated from the measured light intensity. The driving force $F$ is created by a magnetic field gradient that can be applied along either axis of our square box, and we calibrated its magnitude, within 5\%, by 
releasing the gas from the trap in the presence of the force
and measuring the center-of-mass acceleration of the cloud. Changing the direction of the force is equivalent to changing the detection direction, and we use this
symmetry to measure the momentum distribution both parallel and perpendicular to the forcing axis. We tune the scattering length $a$ exploiting a magnetic Feshbach resonance centred at 402.7~G~\cite{Fletcher:2017}. For all our measurements $a$ is in the range $(20-160)~a_0$, where $a_0$ is the Bohr radius.

\subsection{Time-Of-Flight (TOF) measurements}
We take absorption images of our gas after a variable time of ballistic expansion, $t_\mathrm{TOF}$. Just before the expansion, we release the interaction energy by rapidly ($<0.1$~ms) decompressing the out-of-plane confinement. TOF spectra are naturally convolved with the in-trap spatial distribution and are thus not accurate for $k\lesssim mL/(\hbar t_\mathrm{TOF})$. We combine measurements with different $t_\mathrm{TOF}$ \cite{Glidden:2021}, between $8$ and $35$~ms, to extend the range of $k$ values we can reliably probe; the longest $t_\mathrm{TOF}$ minimises finite-size effects at low $k$, while the shortest $t_\mathrm{TOF}$ gives better signal-to-noise ratio at large $k$. For our main TOF measurements (Fig.~1(d), Fig.~3), we use our smaller, $31\mbox{-}\mathrm{\mu m}$ box to limit the low-$k$ range affected by the finite-size effects. We repeated all measurements about 10 times under the same experimental conditions.

\subsection{Principal Component Analysis (PCA)}
PCA extracts orthonormal modes within a data set based on the spectral decomposition of its covariance matrix~\cite{Segal:2010,Dubessy:2014}.
Our data set (for Fig.~2(a)) is a series of {\it in-situ} density profiles obtained by absorption imaging at $J=81$ different times, $t \in [0,90]$~ms, after the initiation of the driving. For each $t$ we repeated the experiment about 5 times and used the 81 averaged profiles to get the 80 PCA modes. In the $b_j(t)$ plots (which have 81 time steps), we filtered out high-frequency noise using a 3-point moving average.

\subsection{Bragg spectroscopy}
The two far-off-resonant laser beams used for Bragg spectroscopy~\cite{Kozuma:1999a,Stenger:1999b} are detuned from each other by a frequency $\Delta \nu$ and have in-plane wavevectors ${\bf k}_1$ and ${\bf k}_2$ such that the recoil momentum $\hbar {\bf k}_{r}=\hbar ({\bf k}_1-{\bf k}_2)$ is aligned with one of the box axes, and $k_r \approx 15~\mathrm{\mu m^{-1}}$ is larger than $\kD$, so the diffracted atoms leave the trap \cite{Gotlibovych:2014}. Measuring the number of diffracted atoms, $N_{\rm diff}$, as a function of $\Delta \nu$ gives the line-integrated distribution parallel to ${\bf k}_r$. The duration of our Bragg pulse is $\tau \approx 1.5$~ms, which gives $k$-space resolution of $\Delta k=2\pi m/(\tau \hbar k_r) \approx 0.17 \ {\rm \mu m^{-1}}$. The absolute values of $N_{\rm diff}$ also depend on the intensity of the Bragg beams, so the measured $\bar{n}_{k, x(y)}$ are not automatically normalised, and in the main text we normalise them to their common value measured for $k=0$ and $t=0$. All measurements were repeated about 20 times under the same experimental conditions.

\subsection{Energy in the isotropic cascade}
To calculate  $E_c(t)$ (Fig.~3(b)) we integrate $\varepsilon_k $ for $k>1/\xi$. Theoretically, the integral could be extended to infinity, but experimentally, integrating to unnecessarily large $k$ (where there is no real atomic population) just adds noise; instead, we cut off the integral at $6\ {\rm \mu m}^{-1}$, which is larger than both the $\kD=5.1(6)\ \mathrm{\mu m}^{-1}$ estimated from $\UD$ and the $\kD$ observed in the saturation of $\kcf$ (Fig.~3(c)). Note that we consider only the increase of the kinetic energy at high $k$ and neglect any changes in the interaction energy. This is a good approximation since for $k>1/\xi$ the excitation energy is mostly kinetic and since in a box trap changing $n_k$ does not change the average gas density, so small changes in the total interaction energy arise only due to modification of the zero-distance second-order correlation function~\cite{Pethick:2002}. Calculating the size of this effect is non-trivial for a far-from-equilibrium gas, but based on an equilibrium calculation, we could be underestimating the rate of growth of the total energy by up to 15\%, which is less than the 20\% systematic uncertainty in $\epsilon$ injected at $\kFv$.
Finally, note that in Fig.~3 at $t\approx t^*$ the fraction of all particles that is at $k>1/\xi$ is about $30\%$, and that for our density and interaction strength (here $\g = 0.018$) the Berezinskii-Kosterlitz-Thouless critical temperature for the transition to superfluidity is $410$~nK \cite{Prokofev:2001,Hadzibabic:2011}, corresponding to $8.5$~kHz, so even the fully developed turbulent state has a comparatively very low energy per particle.

\end{document}